\documentclass[conference]{IEEEtran}
\IEEEoverridecommandlockouts
\usepackage{cite}
\usepackage{amsmath,amssymb,amsfonts}
\usepackage{algorithmic}
\usepackage{graphicx}
\usepackage{textcomp}
\usepackage{xcolor}
\usepackage{url}

\usepackage{selinput}\SelectInputMappings{adieresis={ä},germandbls={ß}}
\usepackage{algorithmic}
\usepackage[ruled,vlined,linesnumbered,commentsnumbered]{algorithm2e}
\algsetup{linenosize=\scriptsize}

\SetCommentSty{mycommfont}
\usepackage{leftidx}
\usepackage{amsthm}
\usepackage[capitalise]{cleveref}
\usepackage{soul}
\usepackage{svg}
\usepackage{subfigure}
\usepackage{array}
\usepackage{multirow}
\usepackage{hhline}
\usepackage{booktabs}
\usepackage[font=footnotesize]{caption}

\newcolumntype{L}[1]{>{\raggedright\let\newline\\\arraybackslash\hspace{0pt}}m{#1}}
\newcolumntype{C}[1]{>{\centering\let\newline\\\arraybackslash\hspace{0pt}}m{#1}}
\newcolumntype{R}[1]{>{\raggedleft\let\newline\\\arraybackslash\hspace{0pt}}m{#1}}

\def\BibTeX{{\rm B\kern-.05em{\sc i\kern-.025em b}\kern-.08em
    T\kern-.1667em\lower.7ex\hbox{E}\kern-.125emX}}

\usepackage[shortcuts,acronym]{glossaries}
\newacronym{aeica}{AeICA}{Adaptive Extraction-based ICA}
\newacronym{auc}{AUC}{Area Under the receiver operating characteristic Curve}

\newacronym{bss}{BSS}{Blind Source Separation}

\newacronym{cdf}{CDF}{Cumulative Distribution Function}
\newacronym{cots}{COTS}{Commercial off-the-shelf}
\newacronym{coin}{COIN}{Computing in the Network}
\newacronym{comnetsemu}{ComNetsEmu}{Communication Networks Emulator}
\newacronym{cs}{CS}{Compressed Sensing}

\newacronym{de}{DE}{Data Extraction}
\newacronym{dr}{DR}{Data Reconstruction}

\newacronym{ica}{ICA}{Independent Component Analysis}
\newacronym{iot}{IoT}{Internet of Things}
\newacronym{iiot}{IIoT}{Industrial Internet of Things}

\newacronym{ml}{ML}{Machine Learning}
\newacronym{mauc}{mAUC}{mean-AUC}
\newacronym{mtu}{MTU}{Maximum Transmission Unit}

\newacronym{nf}{NF}{Network Function}
\newacronym{nfv}{NFV}{Network Function Virtualization}

\newacronym{pica}{pICA}{progressive ICA}
\newacronym{pauc}{pAUC}{partial-AUC}

\newacronym{qos}{QoS}{Quality of Service}

\newacronym{roc}{ROC}{Receiver Operating Characteristic}
\newacronym{rtt}{RTT}{Round-Trip-Time}

\newacronym{sdr}{SDR}{Source-to-Distortion Ratio}
\newacronym{sdn}{SDN}{Software Defined Networking}
\newacronym{sfc}{SFC}{Service Function Chain}
\newacronym{sme}{SME}{Separation Matrix Estimation}
\newacronym{svm}{SVM}{Support-Vector Machine}

\newacronym{urllc}{URLLC}{Ultra-Reliable and Low-Latency Communications}

\newacronym{vnf}{VNF}{Virtualized Network Function}

\begin{document}
\title{In-Network Processing Acoustic Data for Anomaly Detection in Smart Factory}
\author{
	\IEEEauthorblockN{
		Huanzhuo Wu\IEEEauthorrefmark{1},
		Yunbin Shen\IEEEauthorrefmark{1},
		Xun Xiao\IEEEauthorrefmark{3}, Artur Hecker\IEEEauthorrefmark{3} and
		Frank H.P. Fitzek\IEEEauthorrefmark{1}\IEEEauthorrefmark{2}\\
	}
	\IEEEauthorblockA{\IEEEauthorrefmark{1}Deutsche Telekom Chair of Communication Networks - Technische Universität Dresden, Germany
	}
	\IEEEauthorblockA{\IEEEauthorrefmark{2}Centre for Tactile Internet with Human-in-the-Loop (CeTI)}
	\IEEEauthorblockA{\IEEEauthorrefmark{3}Munich Research Center, Huawei Technologies, Muenchen, Germany
	}
	\emph{Email:\{huanzhuo.wu$\mid$frank.fitzek\}@tu-dresden.de, yunbin.shen@mailbox.tu-dresden.de, \{xun.xiao$\mid$artur.hecker\}@huawei.com}

\thanks{This is a preprint of the work~\cite{Wu2021NetworkProcessingAcoustic}, that has been accepted for publication in the proceedings of the 2021 IEEE Global Communications Conference.}
}

\maketitle

\begin{abstract}
Modern manufacturing is now deeply integrating new technologies such as 5G, Internet-of-things (IoT), and cloud/edge computing to shape manufacturing to a new level -- Smart Factory. Autonomic anomaly detection (e.g., malfunctioning machines and hazard situations) in a factory hall is on the list and expects to be realized with massive IoT sensor deployments. In this paper, we consider acoustic data-based anomaly detection, which is widely used in factories because sound information reflects richer internal states while videos cannot; besides, the capital investment of an audio system is more economically friendly. However, a unique challenge of using audio data is that sounds are mixed when collecting thus source data separation is inevitable. A traditional way transfers audio data all to a centralized point for separation. Nevertheless, such a centralized manner (i.e., data transferring and then analyzing) may delay prompt reactions to critical anomalies. We demonstrate that this job can be transformed into an in-network processing scheme and thus further accelerated. Specifically, we propose a progressive processing scheme where data separation jobs are distributed as microservices on intermediate nodes in parallel with data forwarding. Therefore, collected audio data can be separated $43.75\%$ faster with even less total computing resources. This solution is comprehensively evaluated with numerical simulations, compared with benchmark solutions, and results justify its advantages.
\end{abstract}

\begin{IEEEkeywords}
In-network Computing, IoT, Smart Factory, Anomaly Detection
\end{IEEEkeywords}

\section{Introduction}\label{sec:introduction}

The contemporary industry is in the dawn of the 4th revolution towards full digitization and intelligence. Deep integration of emerging technologies such as 5G, Internet-of-things (IoT), artificial intelligence (AI), and cloud computing~\cite{wollschlaeger2017future} is happening right now and brings manufacturing to a new level -- Smart Factory. 
A critical operational task in manufacturing is anomaly detection for machines malfunctioning on the production line. It prevents machines and/or  products from any serious damage and economic losses~\cite{civerchia2017industrial}. In the past, this detection was manually done by on-site inspectors, who were replaced by remote inspectors monitoring on screens. In the future, such inspection would be fully autonomous by analyzing  data from IoT sensors and necessary reactions will be triggered without human intervention.

For anomaly detection, acoustic data can reflect the internal states of machines that are not visible through videos~\cite{ellis1996prediction}. For a long time, acoustic data are used by experienced workers, who can directly judge whether or not a machine works properly by hearing. In addition, audio devices are much cheaper than professional cameras thus more friendly to capital investments. Therefore,  acoustic data-based anomaly detection will still play an important role in future smart factory operations.


Video data is naturally separated when captured with cameras, but this is not the case for acoustic data because sounds interfere with each other by nature. Thus, anomaly detection based on acoustic data is more challenging as original signal data have to be restored first. A natural idea is to first transfer all data to a centralized node; when all data are received, a sort of \ac{bss}~\cite{comon2010handbook} algorithm is applied to separate mixed data. \ac{bss} candidates include \ac{ica}-based methods~\cite{hyvarinen1999fast, pham1997blind, Wu2020} or neural network-based methods~\cite{9367428, luo2019conv}. However, forwarding and then analyzing could delay critical decision-making actions due to $i$) possibly long waiting time of transferring the data, and $ii$) possibly long execution time of running the algorithm on a single node. Clearly, the realization of autonomous anomaly detection requires a better solution.

In this paper, we tackle this problem from another angle: instead of sending all data and then waiting for the separation result, we are thinking if the whole task can be accelerated by distributing the data separation job on intermediate forwarding nodes. In other words, we try to transform the centralized manner into an in-network processing manner to speed up the entire job. The key idea is sketched as follows: a new \emph{lightweight processing logic} is proposed and deployed on every intermediate node running with local spare compute resources as microservices; every node best-effort computes a temporal result, which is a  \emph{solution matrix} that is roughly estimated to restore the original data; this temporal result (i.e., the solution matrix) will be progressively optimized along the forwarding path. Specifically, a modified \ac{ica} algorithm is proposed so that progressive improvements on every node can be maximized in order to fit such a distributed processing scheme; by doing so, at the final destination (i.e., the last hop), an optimal solution matrix shall be ready with sufficient precision approximately. In summary, our key contributions can be summarized as follows:
\begin{enumerate}
		\item We propose an in-network processing solution for acoustic data-based anomaly detection, which is demonstrated as an example how audio data separation can be accelerated up to $43.75\%$ by utilizing intermediate computing resources in networks;
		\item We design a specific processing logic for intermediate nodes with a modified \ac{ica} algorithm, making the acoustic data restoration to be distributedly executable as microservices and yield a faster convergence speed;
		\item We conduct comprehensive simulations and numerical results which justify the effectiveness of our proposed scheme.
\end{enumerate}

To the best of our knowledge, technically, this is the first work that studies how to transform  a \ac{bss} algorithm into an in-network processing scheme, overcoming a key constraint where traditional \ac{bss} can be mainly executed on a centralized node.

The rest of the paper is organized as follows. In~\cref{sec:relatedwork}, a literature review is provided and the main differences of our solution are highlighted; in~\cref{sec:method}, we present full details of our solution. After that, in~\cref{sec:results_analysis}, comprehensive numerical results will be presented, and \cref{sec:conclusion} concludes this paper.

\section{Related Work}\label{sec:relatedwork}


In-network computing/processing -- a joint consideration of computing and communication -- raises increasing research interests because network elements now have much more powerful computing capabilities.
In-network processing empowered with microservices provides a new possibility and  flexibility to support emerging applications that require low latency, high bandwidth, and/or high reliability~\cite{kunze-coin-industrial-use-cases-04}.


Existing in-network processing studies focus more on how to embed computing tasks into a network, deriving the processing results closer to the users~\cite{kunze-coinrg-transport-issues-03}.
For example, the work in \cite{app11052177} decomposes an image detection job as a service function chain. Since the processing can be simply done at closer nodes, it reduces the latency by more than $25\%$. 
Similarly, in~\cite{glebke2019towards} a computer vision task, a real-time line-following service, is decomposed and deployed in a network with a series of programmable network devices.
These works consider only the processing location and network transport issues. On the other hand, the processing logic does not have to be changed or modified.

In contrast, not every task can be deployed in a network straightforwardly. An exception is running a \ac{bss} algorithm for mixture data separation.
This is because the algorithm cannot be simply split into sub-tasks, running on multiple nodes in parallel or a chain. Differently, a joint task is solved by distributed nodes coordinating with each other. Therefore, our problem considers how a \ac{bss} algorithm can be run on distributed nodes. This goal is clearly more sophisticated than pure task decomposition.

When it comes to a \ac{bss} problem, many candidate options are available. One  school is machine learning (ML) based on neural networks (NNs), such as Y-Net~\cite{9367428}, Conv-TasNet~\cite{luo2019conv}. However, for our problem, ML-based solutions are less interesting because $i$) it is hard to obtain enough representative and labeled training data, $ii$) training an NN model is time-consuming and resource-intensive, and $iii$) once deployed on nodes, NN models are inflexible to be updated.
Additionally, to maximize their performance, ML-based solutions require special hardware (e.g., GPUs), which barely appears on a network device. Another school is \ac{ica} algorithms, working directly with input data and require only average hardware capability. Typical \ac{ica} algorithms are FastICA~\cite{hyvarinen1999fast}, InfoMax~\cite{pham1997blind}, and CdICA~\cite{Wu2020}. 
They are free from the constraints of ML-based solutions, so more feasible to fit an in-network processing scheme.

However, existing work only provides centralized \ac{ica} algorithms, which cannot be trivially transplanted to an in-network processing framework. The main reason is: they require all data to calculate a final result (i.e., the solution matrix mentioned before). Therefore, simply executing an \ac{ica} algorithm on every node equals repeating the same procedure by multiple nodes. This does not improve the ultimate convergence.
 
Some recent works realized this issue and looked for a distributed version. 
We note that \ac{aeica}~\cite{Wu2012Adaptive} has the potential to be performed in networks but is quite sensitive to initial parameter configurations, which heavily depend on the prior knowledge of input data. According to reported results, the processing time would rather increase in some cases. Its heterogeneous performances fail to fulfill our goal. We thus only pick it as one of our benchmarks.

In summary, the main differences of this work are that: we study a non-trivial in-network processing problem where the task -- data separation -- cannot be simply decomposed to multiple equivalent sub-tasks; secondly, we fill the gap by converting a traditional \ac{ica} algorithm to fit in a distributed environment and overcome the deficiency.

\section{Our Solution}\label{sec:method}

\subsection{Main Idea}\label{subsec:idea}
According to our observations, the performance of a traditional \ac{ica} algorithm is mainly constrained by the following two factors. 
The first one is the time spent on transferring all data. Before all data are received, the algorithm suspends, with unnecessary waiting time even with perfect network conditions.
The second one is the execution time to wait for a full convergence when running on a centralized node. As we will see in~\cref{subsec:bkg}, the main procedure of an \ac{ica} algorithm is to gradient-descent optimize a solution matrix in iterations. Marginal gains usually become much smaller when approaching the convergence point. In other words, more time is spent at later stages but the improvement is little.

Motivated by these observations, our main idea is to transform this sequential processing manner into an in-network processing manner so that data processing jobs can start early on intermediate nodes in parallel with data forwarding. The achieved acceleration depends on the availability of spare compute resources along the forwarding path. Obviously, the more intermediate compute resources available are, the more accelerations intermediate nodes can contribute.
With this idea, new strategies are introduced as our design principle.

We first introduce a \emph{greedy strategy}, wherein once a node finds that its improvement from gradient-descent gets slower, it stops its local iterations and hands over the temporal result to the next hop. However, simply relaying the temporal results does not bring any acceleration. This leads to the next strategy. 

We further introduce a \emph{growing strategy} on the size of the input dataset to make every node only cache a subset data from the whole dataset but the size of the subset data \emph{progressively} increases on the following nodes. With more input data, this thus guarantees that later nodes can better improve the result. 

Together with the greed strategy, the synthetic effect is that every node takes a temporal solution from its last hop as an input (i.e., a better starting point) and consumes more data to derive an even better solution for the following nodes. This can mitigate the problem of spending too much processing time for little marginal improvements at later iteration stages when running in a centralized manner. Clearly, all these new strategies can be easily realized as microservices on distributed nodes.

Before we start to introduce technical details, basic assumptions are clarified here:
\begin{itemize}
	\item We focus on the algorithmic part and its distributed design in this work. Thus we assume that possible node failures are handled by mechanisms at the network resource layer (i.e., a secondary concern here);
	\item We assume that the network resource allocation happens at earlier stages by management layer; our starting point begins with considering the actual service directly (i.e., in-network processing);
	\item We assume that an intermediate node is not as powerful as a \ac{cots} server machine and a single network device cannot afford the whole processing task. However, collectively, the sum of all distributed compute resources is enough.
\end{itemize}


\subsection{Scenario and System Architecture}
\begin{figure}[t!]
	\centerline{\includegraphics[width=\columnwidth]{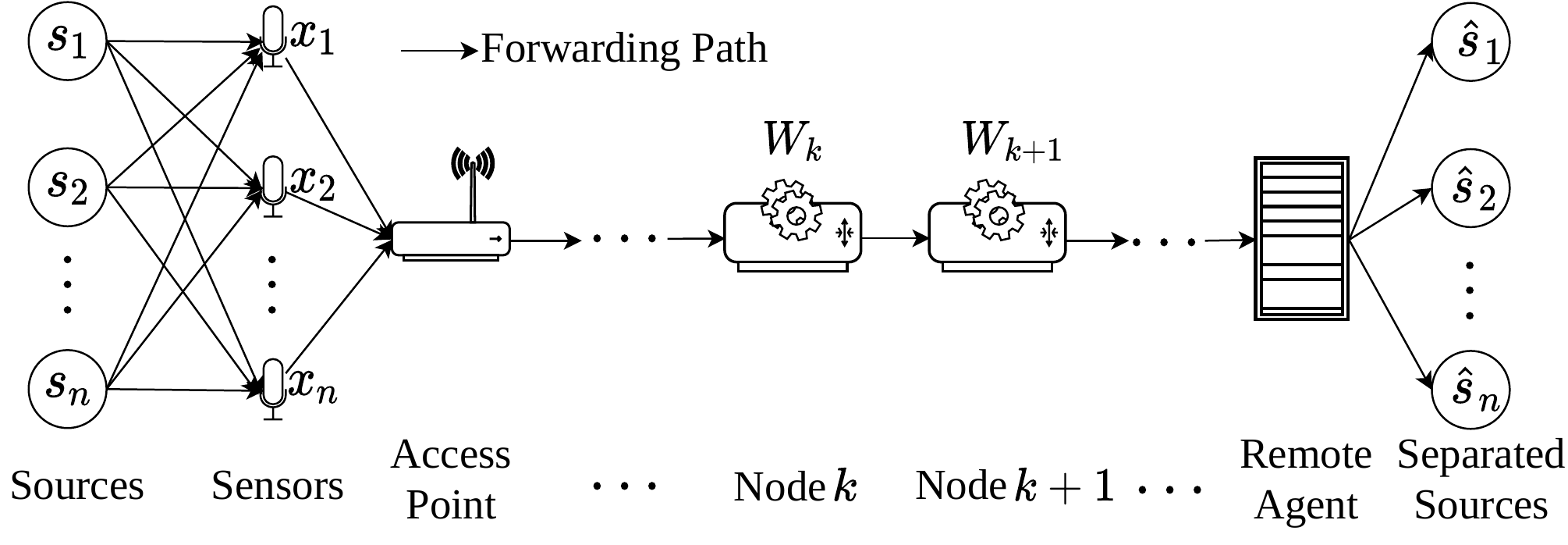}}
	\caption{An in-network processing architecture.
	}
	\label{fig:network_topology}
\end{figure}

An example scenario considered here is illustrated in~\cref{fig:network_topology}. Specifically, every IoT sensor collects sound data from the machine ``source'' it attaches to. One audio sound signal interferes in the air with sounds from other machines (as noise). Every IoT sensor constantly sends collected data to a wireless access point (AP). The AP node can be either a WiFi AP or a cellular base station of a (non-public) network. This AP keeps forwarding the data (from all sensors) to the backend ``Remote Agent'' over a forwarding path. This path can be either dynamically determined based on a routing protocol or statically configured. The forwarding path consists of some intermediate ``Node $k$'' and ``Node $k+1$''.


Given this system, there are $n$ working machines indexed by $i$. The original data denoted by $s_i$ is generated from the $i$-th working machine. As said, the original signal data $s_i$ will be distorted to imperfect data $x_i$ due to  mutual interference. With $m$ time slots, collected data at the AP aggregate to a data matrix $X$. Accordingly, the original counterpart of $X$ is thus a source matrix $S$. We formulate this mutual interference effect as the operation in~\cref{eq:bss}: \begin{equation}\label{eq:bss}
	X = A\times S
\end{equation}
where distorting the original source data $S$ to $X$ is modeled by a mixing matrix $A$. Mathematically, a \ac{bss} problem is an inverse operation of~\cref{eq:bss}: $\hat{S} = A^{-1} \times X = W \times X$,
where original data $\hat{S}$ are estimated by deriving the solution matrix $W$, which will be applied on the input data $X$ for data restoration.

Our in-network processing scheme introduces a new \emph{processing logic} as a microservice running on every intermediate node (gear icons in~\cref{fig:network_topology}). As introduced in the previous section, while forwarding the collected data $X$, node $k$ will start processing with a subset data sampled from $X$. This aims to roughly but quickly calculate a temporal solution matrix $W_k$, then passes $W_k$ to node $k+1$ (recall the greedy strategy). Node $k+1$ will evolve the temporal solution to a better solution matrix $W_{k+1}$ with a larger size of subset data (recall the growing strategy). We will see that such a progressive manner yields a faster converging speed without sacrificing the precision on the final solution matrix $W$. 


Next, we detail the new processing logic, namely \acrfull{pica}, which functions as the key component calculating the solution matrix $W$.
%
Besides, all notations are summarized in~\cref{tab:notation}.
\begin{table}[t!]
    \def\arraystretch{1.2}
    \centering
    \caption{Notation}
    \label{tab:notation}
    \begin{tabular}{C{1.1cm}|L{6.7cm}}
    \toprule[1pt]
    \bf Notation & \bf Meaning\\
    \midrule[1pt]
    $s_i$ 	& Original source data from the $i$-th machine. \\
    $S$ 	& Original data matrix from all machines. \\    
    $x_i$ 	& Data collected by the $i$-th IoT sensor. \\
    $X$ 	& Data matrix received by AP. \\
    $m$ 	& Total time span. \\
    $A$ 	& Mixing matrix mimic mutual interference. \\
    $W$ 	& Solution matrix separating $S$ out of $X$. \\
    $W_k$ 	& Solution matrix on the $k$-th network node. \\
    $\hat{s}_i$ & The $i$-th separated source signal. \\
    $\hat{S}$ 	& Separated source matrix consisting all $\hat{s}_i$. \\
    \midrule[0.5pt]
    $Tol$ 	& Threshold precision of convergence. \\
    $\mu_k$ 	& Sampling step for $X$ on node $k$. \\
    $\alpha_k$ 	& Parameter to denser the sampling from $X$ on node $k$. \\
    $g_k$ 		& Gradient of Newton's iteration on node $k$. \\
    $\hat{h}_g$	& Threshold of gradient to Newton's iteration on node $k$. \\
    $\Delta_l$ & Difference between two consecutive iterations\\
    \bottomrule[1pt]
    \end{tabular}
\end{table}

%
%
%
%
%
%
%
%

\subsection{Intermediate Node Processing Logic}

\subsubsection{Preliminary}\label{subsec:bkg}

A general \ac{bss} problem blindly recovers original data from mixture data without prior knowledge. One of the popular methods is \ac{ica}~\cite{comon2010handbook} estimating the solution matrix $W$ mentioned above. If original data is $i$) statistically independent and $ii$) non-Gaussian distributed, conventional ICA-based algorithms (e.g., FastICA~\cite{hyvarinen1999fast}) maximize the non-gaussianity of observed data to separate the mixtures.
This can be done with an approximation of Newton's Iteration $\mathcal{N}$:
\begin{equation}\label{eq:newton_def}
	(W_{l}, \Delta_l) = \mathcal{N}(X, W_{l-1}),~l \in \mathbb{N},
\end{equation}
where $X$ is the given mixture data, $W_{l}$ is the estimation of separation matrix in the $l$-th iteration.
$\Delta_l$ is the difference between two consecutive iterations, defined by:
\begin{equation}\label{eq:diff_def}
	\Delta_l = W_{l} \times W_{l-1}^T - I,~l \in \mathbb{N}, 
\end{equation}
where $I$ is an identity matrix. Newton's Iterations keep updating $W_l$ until it converges.
It sets a desired tolerance $Tol$ indicating one of the stoppage criteria (e.g., $\Delta_l \leq Tol$) to exit the iterations.
%
The original ICA above requires all data available on a single point, without considering the issue of running in a distributed environment. 

Based on the classical version, ICA variants are proposed. For example, \ac{aeica} is introduced in~\cite{Wu2012Adaptive}, where a very basic growing size of subsets of data is used to calculate the solution matrix $W$ in iterations.
However, the performance of \ac{aeica} is heterogeneous case-by-case, because the parameter controlling the subset data sampling distance is chosen based on an ideal assumption where the distribution of the data features is uniform. This assumption does not hold in reality because the prior knowledge on $X$ is usually unknown. This means that the sampled subset data are not always representative, which leads to inconsistent separation results.


\subsubsection{Our \ac{pica}}\label{subsec:PICA}
\ac{pica} runs as an in-network microservice. On an intermediate node $k$, $i$) it keeps forwarding the data to the next node $k+1$; and $ii$) it starts with the temporal result (i.e., a solution matrix $W_{k-1}$) provided by node $k-1$ and further improves the result for the next node $k+1$. 

Different to \ac{aeica}, \ac{pica} makes two substantial modifications. The first one is a new sampling strategy to sample subset data, which eliminates the dependence on the prior knowledge on input data $X$; and the second one is a new set of stoppage criteria where node $k$ uses to judge whether or not local iterations should continue. 

Now we introduce the first modification -- the new sampling strategy. Our \ac{pica} introduces a controlling parameter $\alpha_{k}$, instead of assuming any prior knowledge as in AeICA. Every node will dynamically adjust $\alpha_{k}$ in order to control the variety of the sampled subset data based on the outcome from the last hop.
Specifically, a node $k$ can tune the value of $\alpha_k$ so that the sampling step $\mu_k$ becomes $\alpha_k$ times smaller than the step value used in the last hop:
\begin{equation}
	\mu_k \leftarrow \frac{\mu_{k-1}}{\alpha_k}~,	
\end{equation}
The decreasing value of $\mu_k$ leads to the size of sampled data sequentially increasing on consecutive nodes. For example, $\alpha_k = 2$ means on every hop the sampled data are doubled than that of the previous hop, since the sampling step is halved .

%

With the sampled subset data, denoted as ${}_{\mu_k}X$, instead of re-computing a solution matrix $W_k$, node $k$ continues with the temporal result $W_{k-1}$ provided by the last hop:
\begin{equation}
	(W_k,\Delta_k) \leftarrow \mathcal{N}({}_{\mu_k}X, W_{k-1})~,
\end{equation}
where stoppage criteria of a local Newton's Iteration, which is the second modification to \ac{aeica}, are as follows.

The first criterion is the local convergence tolerance $Tol$, which characterizes the required precision of solution matrix $W_k$. Obviously, if $W_k$ is seen to arrive at the required precision (i.e., $\Delta_k\leq Tol$), the iteration on node $k$ can stop. Note that this does not mean $W_k$ is globally optimal because it is calculated based on a local subset of data sampled with $\mu_{k}$. On node $k+1$, the size of the subset of data will increase (e.g., $\alpha_{k + 1} = \alpha_k \times 2$), so $W_k$ will be further improved if possible.

%

Another criterion is to indicate whether or not the local marginal gain becomes too small (recall our greedy strategy). Node $k$ continues only if the gradient value $g_k$ is still large enough in iterations. 
If the current $g_k$ appears too small ($g_k\leq \hat{h}_g$), further iterations will not improve the solution matrix $W_k$ significantly anymore. Recall the existing \ac{aeica}, its stoppage criterion however only employs a sole convergence tolerance based on $\mu_{k}$ without measuring the marginal gain. This leads \ac{aeica} to waste execution time on pursuing little improvements on one node. 

After node $k$ exits its iteration, it will relay its solution matrix $W_k$ together with the stepping parameter $\mu_{k}$ to node $k+1$.
If a node sees $\mu_k$ diminishes to smaller than $1$, which means that all data have been used on the previous node. Then this node knows that it is the last stop of the progressive procedure. The action is to run a classical \ac{ica} until the precision of the final solution matrix $W$ satisfies the predefined $Tol$. 
Note that most of the jobs for optimizing $W$ have been done on previous nodes, thus the last step only requires little extra effort. As a final output, the solution matrix $W$ is applied on the input data $X$ to estimate original data $\hat{S}$, which will be eventually used for anomaly detection. The processing logic of node $k$ (i.e., the microservice's template) is summarized in~\cref{alg:PICA}\footnote{The algorithm is open sourced at \url{https://github.com/Huanzhuo/pICA}.}.

\setlength{\textfloatsep}{0pt}
\begin{algorithm}[t!]
\caption{pICA algorithm on node $k$.}\label{alg:PICA}
\SetAlgoLined
\SetKwInOut{Input}{input}\SetKwInOut{Output}{output}\SetKwInOut{Initialize}{initialize}
 \Input{$X \in \mathbb{R}^{n \times m}$, $W_{k-1} \in \mathbb{R}^{n \times n}$, $\mu_{k-1}$, $\alpha_{k}$.}
 \Output{$W_k \in \mathbb{R}^{n \times n}$, $\mu_k$, $\alpha_{k+1}$ or $\hat{S}$}
 Update stepping parameter $\mu_k \leftarrow \frac{\mu_{k-1}}{\alpha_k}$\;
 Sample subset of sensing data ${}_{\mu_k}X$\;
\While{True}{
    $W_k \leftarrow \mathcal{N}({}_{\mu_k}X, W_{k-1})$ \;
    \eIf(\tcp*[f]{Intermediate node of computing.}){$\mu_k>1$}
    {
        \uIf(\tcp*[f]{Reach tolerance.}){$\Delta_k<Tol$}
        {
            $\alpha_{k+1} = \alpha_k \times 2$;
            break\;
        }
        \uElseIf(\tcp*[f]{Small gradient variation.}){$g_k<\hat{h}_g$}
        {
            $\alpha_{k + 1} = max(2, \frac{\alpha_k}{2})$;
            break\;
        }
        $k++$\;
    }(\tcp*[f]{Last node of computing.})
    { 
        \If{$\Delta_k<Tol$}
        {
            $\hat{S} = W_k \times X$;
            break\;
        }
    }
}
\end{algorithm}
\section{Numerical Results} \label{sec:results_analysis}
\subsection{Simulation Setup}\label{sec:experimental_design}

\subsubsection{Dataset}\label{subsec:evaluation_dataset}
We pick a published data set from~\cite{Purohit_DCASE2019_01}, called MIMII\footnote{MIMII: Malfunctioning Industrial Machine Investigation and Inspection} for evaluation. It collects normal and anomalous operating sound data of  $n=4$ types of machines (including valves, pumps, fans, and slide rails). Every segment is a $10$-second audio (single-channel and sample rate is $16$kHz). The size of one data source $s_i$ is $m = 160k$ (sample rate$\times$duration). Since we have $4$ types of data sources, the original data $S$ is a $4 \times 160k$ matrix.

A $4\times 4$ mixing matrix $A$ generated from a standard distribution will be applied to the original data matrix $S$ according to~\cref{eq:bss} to simulate the effect of mutual interference. At the AP node, it will receive the data matrix $X$ with the same size as $S$. 

Note that this is just $10$s audio data of $4$ types of machines. In reality, there will be much more machine types and much longer audio data, time of data transferring and after that processing them may delay critical anomaly detection.


\subsubsection{Scenarios} Our simulation considers five network configurations: $k$ intermediate nodes ($k = 0, 3, 7, 10$ and $15$, respectively), representing a low to a high level of resource availability.
For each given number of intermediate nodes $k$, we run our simulation $50$ times to exhibit the randomness of the mixing matrix $A$.

We implemented the proposed \ac{pica} in Python, which can be directly deployed as microservices. Multi-node network configurations are virtualized on a \ac{cots} server with an i7-6700T CPU with 16GB RAM using Ubuntu 18.04 LTS.

For comparison, we choose FastICA and \ac{aeica} discussed in~\cref{sec:relatedwork}. Note that FastICA can only be executed on a single node. When $k=0$ (i.e., no intermediate node available), actually all candidates (FastICA, AeICA, and pICA) run in a centralized manner. 

\subsubsection{Measured Metrics}\label{subsec:evaluation_metrics}
The first metric is total \emph{ processing time} $t_p$. We use a python module \texttt{time} (with $1$ microsecond ($\mu s$) precision). $t_p$ covers the entire data processing time in our network system, i.e., from the moment the data enters the network to the end of the estimated original data $\hat{S}$ derived.

The second metric is \emph{precision}. A \ac{sdr} metric from~\cite{vincent2006performance} is used to quantify the precision of the estimated original data $\hat{S}$. Its definition is given in~\cref{eq:sdr}:
\begin{equation}\label{eq:sdr}
    \ac{sdr} = 10\cdot\log_{10} \frac{\|s_{truth}\|^2}{\|e_{interf} + e_{noise} + e_{artif}\|^2}~,
\end{equation}
where $s_{truth}$ is the known truth from the picked dataset, $e_{interf}$, $e_{noise}$, and $e_{artif}$ are the respective errors of interference, noise, and artifacts errors, which can be evaluated with the output of any separation algorithm (like our $\hat{S}$) with an open-source BSS Eval Toolbox~\cite{stoter20182018}.
\ac{sdr} is most widely used metric nowadays because different types of errors are comprehensively considered.

\subsection{Processing Time and Precision}\label{sec:latency_accuracy}
\begin{figure}[htb!]
\centerline{\includegraphics[width=\columnwidth]{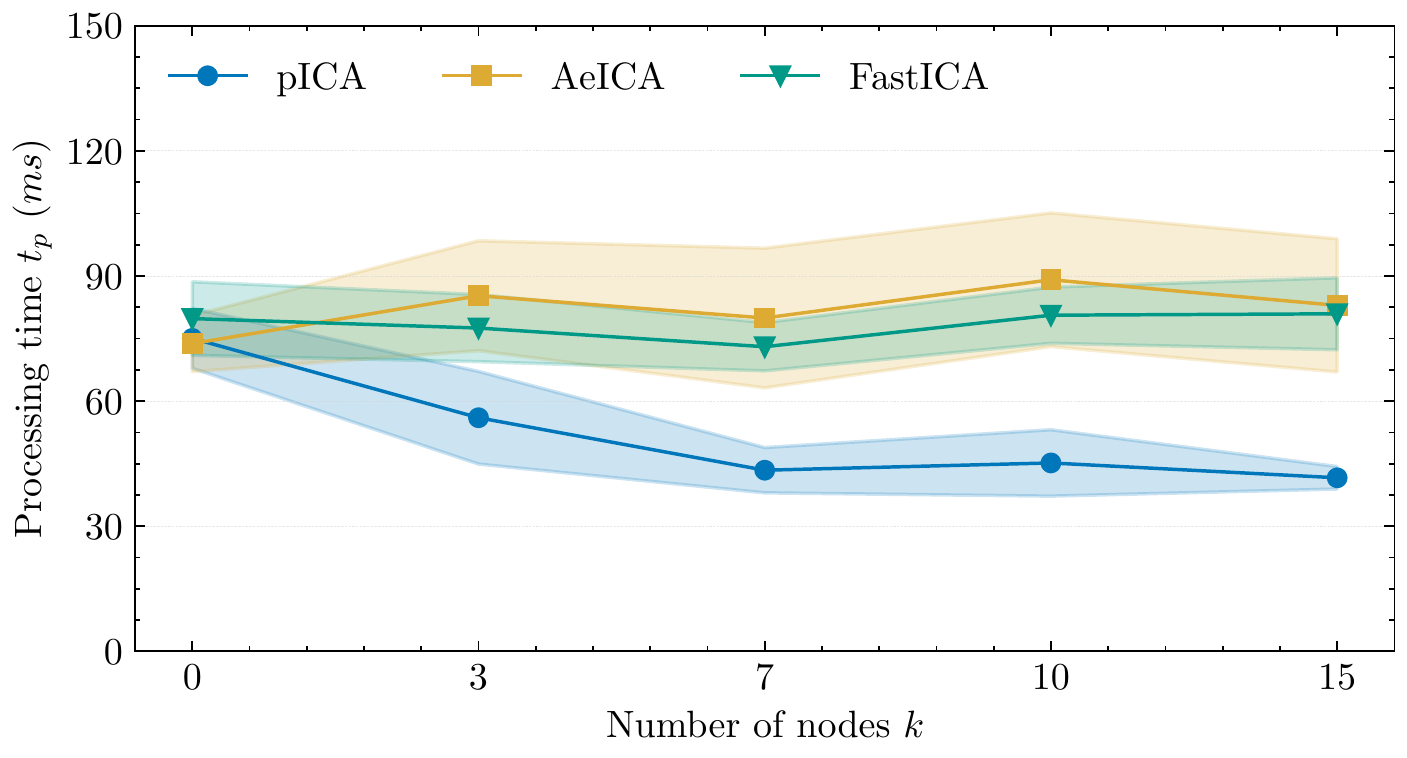}}
\caption{Processing time comparison, $\mu_0 = 4130, \alpha_0=2$.}
\label{fig:time_on_nodes}
\end{figure}

\cref{fig:time_on_nodes} provides a comparison of the processing time of selected algorithms.
We can observe that \ac{pica} gains much more speed-ups with increasing numbers of intermediate nodes $k$. The processing time of \ac{pica} decreases from ca. $80ms$ to ca. $45ms$ ($43.75\%$ faster). Specifically, when $k=0$, non-surprisingly, the three algorithms show similar performances because no acceleration with intermediate nodes. When $k=3$, \ac{pica} starts outperforming the other two. The gap becomes larger when $k=7$ increasing to $k=15$. This proves that \ac{pica} can accelerate more with intermediate resources compared with others. Additionally, such gains tend to be deterministic (i.e., smaller variances) while the other two candidates do not show good stability.
  

\begin{figure}[htb!]
\centerline{\includegraphics[width=\columnwidth]{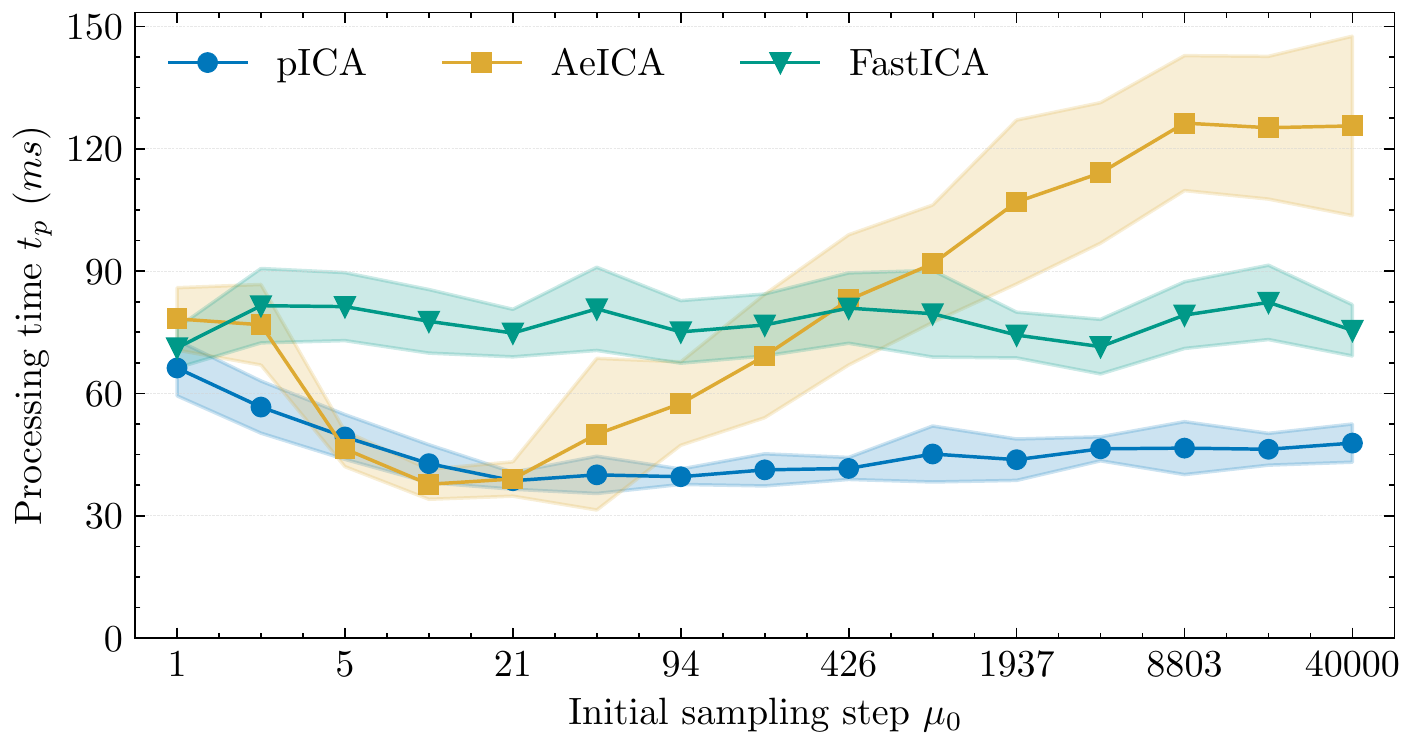}}
\caption{Processing time with different initial sampling steps, $k = 15, \alpha_0=2$.}
\label{fig:time_diff_mu0}
\end{figure}
\cref{fig:time_diff_mu0} further presents the impact of how the sampling step distance $\mu_0$ on processing time. The result shows that our \ac{pica} even prefers starting with a small subset of data (i.e., larger values of $\mu_0$). For example, when $\mu_0=1$ (meaning with all data), \ac{pica} does not yield any acceleration at all; however,  
with $\mu_0=3$, the processing time declines from $65ms$ to $53ms$; with $\mu_0 > 21$, the processing time constantly reduces to around $45$ms. On the other hand, since \ac{aeica} is very sensitive to the parameter $\mu_0$, which depends on the prior knowledge of the input data, its performance becomes worse. Thus, it cannot always benefit from available intermediate nodes. This  justifies our proposed growing strategy.

\begin{figure}[htb!]
\centerline{\includegraphics[width=\columnwidth]{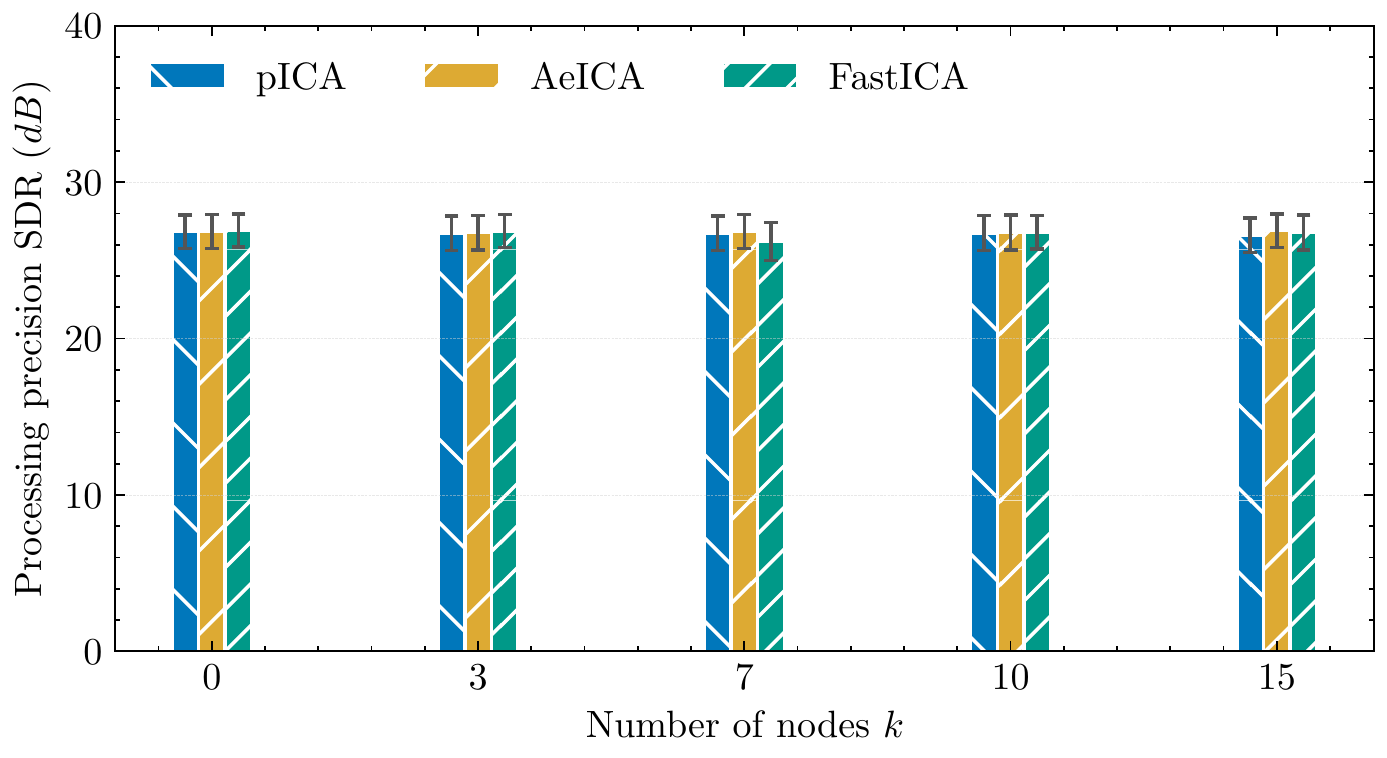}}
\caption{Processing precision comparison, $\mu_0=511, \alpha_0=2$.}
\label{fig:sdr_on_nodes}
\end{figure}
\cref{fig:sdr_on_nodes} compares the achieved precision (\ac{sdr}) with different methods. It proves that our \ac{pica} does not compromise its precision for acceleration but yields an equivalent precision as the other two methods to restore the original data. It again justifies the benefits of the progressive and greedy strategies when \ac{pica} is executed hop-by-hop.


\subsection{Performance of Intermediate Nodes}\label{sec:pica_steps}

\begin{figure}[t!]
\centerline{\includegraphics[width=\columnwidth]{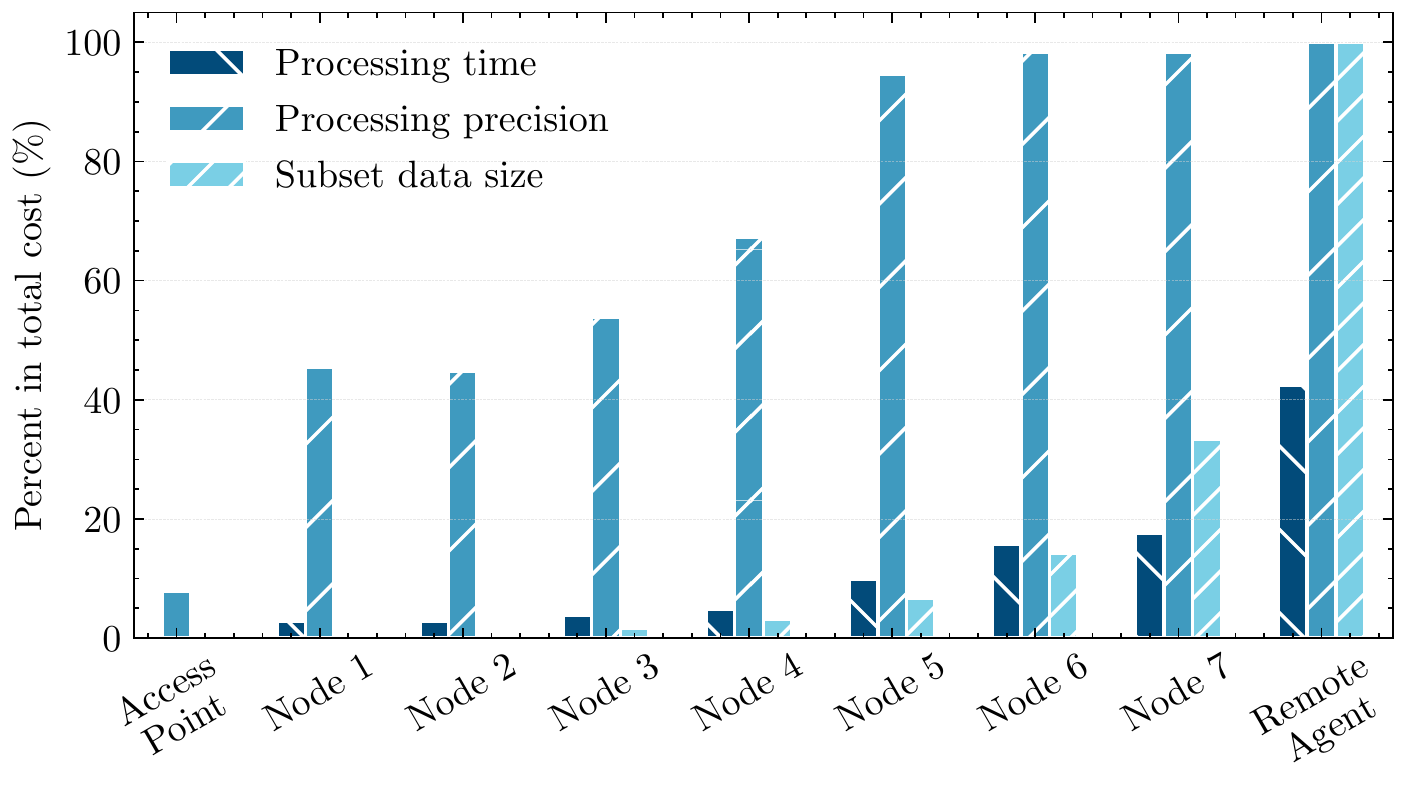}}
\caption{Intermediate execution costs of 
	\ac{pica}, $k = 7$ and $\mu_0 = 500, \alpha_0=2$.}
\label{fig:node_performance}
\end{figure}
At last, we reveal the intermediate procedure by showing the execution costs of \ac{pica} (with $k=7$ and $\mu_0=500$) on individual nodes. 
The result shows that $60\%$ of processing tasks are finished on intermediate nodes and only $40\%$ jobs left to the last node. Meanwhile, the precision of the estimated original data $\hat{S}$ improves quickly ($>60\%$ after node $4$'s processing) with the size of the subset data growing. It reveals that with small amounts of data, a solution matrix with relatively high precision can be derived. This again justifies the effectiveness of our strategy of introducing the only subset of data on every hop.


\section{Conclusion}\label{sec:conclusion}

This paper provides a novel in-network processing solution suitable with microservices for acoustic data separation for anomaly detection: \acrlong{pica}. It is built with in-network nodes and results prove that it indeed accelerates the processing time of mixture data separation by up to $43.75\%$. This gives a new representative use case demonstrating the advantage of in-network processing and how spare resources can be better utilized for critical tasks.


\section*{Acknowledgment}
This work was funded by the Federal Ministry of Education and Research of Germany (Software Campus Net-BliSS 01IS17044), and by the German Research Foundation (Deutsche Forschungsgemeinschaft) as part of Germany’s Excellence Strategy (EXC 2050/1 – Project ID 390696704) Cluster of Excellence “Centre for Tactile Internet with Human-in-the-Loop” (CeTI) of Technische Universität Dresden.

\bibliographystyle{IEEEtran}
\bibliography{bibtex}

\begin{thebibliography}{10}
\providecommand{\url}[1]{#1}
\csname url@samestyle\endcsname
\providecommand{\newblock}{\relax}
\providecommand{\bibinfo}[2]{#2}
\providecommand{\BIBentrySTDinterwordspacing}{\spaceskip=0pt\relax}
\providecommand{\BIBentryALTinterwordstretchfactor}{4}
\providecommand{\BIBentryALTinterwordspacing}{\spaceskip=\fontdimen2\font plus
\BIBentryALTinterwordstretchfactor\fontdimen3\font minus
  \fontdimen4\font\relax}
\providecommand{\BIBforeignlanguage}[2]{{%
\expandafter\ifx\csname l@#1\endcsname\relax
\typeout{** WARNING: IEEEtran.bst: No hyphenation pattern has been}%
\typeout{** loaded for the language `#1'. Using the pattern for}%
\typeout{** the default language instead.}%
\else
\language=\csname l@#1\endcsname
\fi
#2}}
\providecommand{\BIBdecl}{\relax}
\BIBdecl

\bibitem{Wu2021NetworkProcessingAcoustic}
H.~Wu, Y.~Shen, X.~Xiao, A.~Hecker, and F.~H. Fitzek, ``{In-Network} processing
  acoustic data for anomaly detection in smart factory,'' in \emph{2021 IEEE
  Global Communications Conference: IoT and Sensor Networks (Globecom2021
  IoTSN)}.

\bibitem{wollschlaeger2017future}
M.~Wollschlaeger, T.~Sauter, and J.~Jasperneite, ``The future of industrial
  communication: Automation networks in the era of the internet of things and
  industry 4.0,'' \emph{IEEE Industrial Electronics Magazine}, vol.~11, no.~1,
  pp. 17--27, 2017.

\bibitem{civerchia2017industrial}
F.~Civerchia, S.~Bocchino, C.~Salvadori, E.~Rossi, L.~Maggiani, and
  M.~Petracca, ``Industrial internet of things monitoring solution for advanced
  predictive maintenance applications,'' \emph{Journal of Industrial
  Information Integration}, vol.~7, pp. 4--12, 2017.

\bibitem{ellis1996prediction}
D.~P. Ellis, ``Prediction-driven computational auditory scene analysis,'' Ph.D.
  dissertation, Columbia University, 1996.

\bibitem{comon2010handbook}
P.~Comon and C.~Jutten, \emph{Handbook of Blind Source Separation: Independent
  component analysis and applications}.\hskip 1em plus 0.5em minus 0.4em\relax
  Academic press, 2010.

\bibitem{hyvarinen1999fast}
A.~Hyvarinen, ``Fast and robust fixed-point algorithms for independent
  component analysis,'' \emph{IEEE transactions on Neural Networks}, vol.~10,
  no.~3, pp. 626--634, 1999.

\bibitem{pham1997blind}
D.~T. Pham and P.~Garat, ``Blind separation of mixture of independent sources
  through a quasi-maximum likelihood approach,'' \emph{IEEE transactions on
  Signal Processing}, vol.~45, no.~7, pp. 1712--1725, 1997.

\bibitem{Wu2020}
H.~{Wu}, Y.~{Shen}, J.~{Zhang}, I.~A. {Tsokalo}, H.~{Salah}, and F.~H.~P.
  {Fitzek}, ``Component-dependent independent component analysis for
  time-sensitive applications,'' in \emph{2020 IEEE International Conference on
  Communications (ICC)}.\hskip 1em plus 0.5em minus 0.4em\relax Dublin,
  Ireland: IEEE, 2020.

\bibitem{9367428}
H.~{Wu}, J.~{He}, M.~{Tömösközi}, and F.~H.~P. {Fitzek}, ``Y-net: A dual
  path model for high accuracy blind source separation,'' in \emph{2020 IEEE
  Globecom Workshops (GC Wkshps)}, 2020, pp. 1--6.

\bibitem{luo2019conv}
Y.~Luo and N.~Mesgarani, ``Conv-tasnet: Surpassing ideal time--frequency
  magnitude masking for speech separation,'' \emph{IEEE/ACM transactions on
  audio, speech, and language processing}, vol.~27, no.~8, pp. 1256--1266,
  2019.

\bibitem{kunze-coin-industrial-use-cases-04}
\BIBentryALTinterwordspacing
I.~Kunze, K.~Wehrle, and D.~Trossen, ``{Use Cases for In-Network Computing},''
  Internet Engineering Task Force, Internet-Draft
  draft-kunze-coin-industrial-use-cases-04, Nov. 2020, work in Progress
  (Accessed 2021-01-28). [Online]. Available:
  \url{https://datatracker.ietf.org/doc/html/draft-kunze-coin-industrial-use-cases-04}
\BIBentrySTDinterwordspacing

\bibitem{kunze-coinrg-transport-issues-03}
\BIBentryALTinterwordspacing
------, ``{Transport Protocol Issues of In-Network Computing Systems},''
  Internet Engineering Task Force, Internet-Draft
  draft-kunze-coinrg-transport-issues-03, Nov. 2020, work in Progress (Accessed
  2021-01-28). [Online]. Available:
  \url{https://datatracker.ietf.org/doc/html/draft-kunze-coinrg-transport-issues-03}
\BIBentrySTDinterwordspacing

\bibitem{app11052177}
\BIBentryALTinterwordspacing
Z.~{Xiang}, P.~{Seeling}, and F.~H.~P. {Fitzek}, ``You only look once, but
  compute twice: Service function chaining for low-latency object detection in
  softwarized networks,'' \emph{Applied Sciences}, vol.~11, no.~5, 2021.
  [Online]. Available: \url{https://www.mdpi.com/2076-3417/11/5/2177}
\BIBentrySTDinterwordspacing

\bibitem{glebke2019towards}
R.~Glebke, J.~Krude, I.~Kunze, J.~R{\"u}th, F.~Senger, and K.~Wehrle, ``Towards
  executing computer vision functionality on programmable network devices,'' in
  \emph{Proceedings of the 1st ACM CoNEXT Workshop on Emerging in-Network
  Computing Paradigms}, 2019, pp. 15--20.

\bibitem{Wu2012Adaptive}
H.~Wu, Y.~Shen, J.~Zhang, H.~Salah, I.~A. Tsokalo, and F.~H. Fitzek, ``Adaptive
  {Extraction-Based} independent component analysis for {Time-Sensitive}
  applications,'' in \emph{2020 IEEE Global Communications Conference: Selected
  Areas in Communications: Internet of Things and Smart Connected Communities
  (Globecom2020 SAC IoTSCC)}, Taipei, Taiwan, Dec. 2020.

\bibitem{Purohit_DCASE2019_01}
H.~Purohit, R.~Tanabe, T.~Ichige, T.~Endo, Y.~Nikaido, K.~Suefusa, and
  Y.~Kawaguchi, ``{MIMII Dataset}: Sound dataset for malfunctioning industrial
  machine investigation and inspection,'' in \emph{Proceedings of the Detection
  and Classification of Acoustic Scenes and Events 2019 Workshop
  ({DCASE2019})}, November 2019, pp. 209--213.

\bibitem{vincent2006performance}
E.~Vincent, R.~Gribonval, and C.~F{\'e}votte, ``Performance measurement in
  blind audio source separation,'' \emph{IEEE transactions on audio, speech,
  and language processing}, vol.~14, no.~4, pp. 1462--1469, 2006.

\bibitem{stoter20182018}
F.-R. St{\"o}ter, A.~Liutkus, and N.~Ito, ``The 2018 signal separation
  evaluation campaign,'' in \emph{International Conference on Latent Variable
  Analysis and Signal Separation}.\hskip 1em plus 0.5em minus 0.4em\relax
  Springer, 2018, pp. 293--305.

\end{thebibliography}

\end{document}